\documentstyle[12pt,aaspp4,epsf]{article}

\def\etal{{\it et al.\ }}
\def\fun#1#2{\lower3.6pt\vbox{\baselineskip0pt\lineskip.9pt
  \ialign{$\mathsurround=0pt#1\hfil##\hfil$\crcr#2\crcr\sim\crcr}}}
\newcommand{\beq}{\begin{equation}}
\newcommand{\eeq}{\end{equation}}
\newcommand{\gray}{$\gamma$-ray}

\begin{document}

%\date{%\today}

\title{Photodisintegration of Ultrahigh Energy Cosmic Rays: A New Determination}

\author{F. W. Stecker}

\affil{Laboratory for High Energy Astrophysics, Code 661,\\
NASA/Goddard Space Flight Center, Greenbelt, MD 20771, USA.}

\authoremail{stecker@lheapop.gsfc.nasa.gov}

\author{M. H. Salamon}

\affil{Physics Department, University of Utah, Salt Lake City, UT 84112}

\authoremail{salamon@cosmic.physics.utah.edu}

\author{{\bf{\it Astrophys.J.}, 512, in press, 20 February 1999.}}   

\vspace{0.8cm}

%\maketitle

\begin{abstract}

We present the results of a new calculation of the photodisintegration of 
ultrahigh energy cosmic-ray (UHCR) nuclei in intergalactic
space. The critical interactions for energy loss and photodisintegration
of UHCR nuclei occur with photons of the 2.73 K cosmic background radiation 
(CBR) and with photons of the infrared background radiation (IBR). 
We have reexamined this problem making use of a new determination of the IBR
based on empirical data, primarily from IRAS galaxies, consistent with
direct measurements and upper limits from TeV $\gamma$-ray observations.
We have also improved the calculation by  
including the specific threshold energies for the various photodisintegration 
interactions in our Monte Carlo calculation. With the new smaller IBR 
flux, the steepness of the Wien side of the now relatively 
more important CBR makes their inclusion essential for more accurate results.
Our results indicate a significant increase in the propagation time of UHCR
nuclei of a given energy over previous results. We discuss the possible 
significance of this for UHCR origin theory.

\end{abstract}


\section{Introduction}

Shortly after the discovery of the cosmic microwave background radiation (CBR),
it was shown that cosmic ray protons above $\sim$60 EeV ($6 \times 10^{19}$eV) 
should be attenuated by photomeson interactions with CBR photons 
(\cite{gre66,zat66,ste68}).
It was later calculated that heavier cosmic ray nuclei with similar 
{\it total} energies would also be attenuated, but by a different process, 
{\it viz.}, photodisintegration interactions with IBR photons (\cite{pug76}),
hereafter designated PSB).\footnotemark{}\footnotetext{For earlier work and 
other considerations involving the photodisintegration of ultrahigh energy
cosmic-ray nuclei, see Stecker (1969), Tkaczyk, Wdowczyk \& Wolfendale (1975),
and Karakula \& Tkaczyk (1993).}
We will refer to such cosmic rays of total energies above 10 EeV as ultrahigh
energy cosmic rays (UHCR).
 
In the more conventional scenario, UHCRs are charged particles which
must be accelerated to ultrahigh energies by electromagnetic processes at 
extragalactic sites, both because there are no known sites in our galaxy 
which can accelerate and magnetically contain them and also because
most of the observed UHCR air shower events arrive from directions outside of 
the galactic plane. Although such acceleration of charged particles to 
energies above 100 EeV in cosmic sources pushes our present 
theoretical ideas to their extreme, it has been suggested that it may occur 
in hot spots in the lobes of radio galaxies (\cite{bie87,tak90}).

The detection of the two highest energy air shower events yet observed, 
with energies of $\sim 200$ (between 170 and 260) EeV
(\cite{hay94}) and $320\pm 90$ EeV 
(\cite{bir95}) has aggravated both
the acceleration and propagation problems for cosmic-ray physicists. (Very
recently, the AGASA group has presented a total of 6 events of energies
between $\sim 100$ and $\sim 200$ EeV, including the one cited above,
observed since 1990 (\cite{tak98}).)
How does nature accelerate particles to these extreme energies and how do they
get here from extragalactic sources (\cite{elb95})?
To answer these questions, 
new physics has been invoked, physics involving the formation and 
annihilation of topological defects (TDs) which may 
have been produced in the very earliest 
stages of the big bang, perhaps as a result of grand unification. A TD
annihilation or decay scenario has unique observational consequences, 
such as the copious 
production of UHCR neutrinos and $\gamma$-rays (\cite{sig96} and refs. therein;
\cite{bha98}).
A new ground-based detector array experiment named after Pierre Auger 
(\cite{cro92}) and an interesting satellite experiment called {\it OWL} 
(\cite{orm97}) have been proposed to test look for such
consequences. 

\section{Propagation of UHCR Nuclei}

A UHCR {\it proton} of energy $\sim$ 200 EeV has a lifetime against 
photomeson losses of $\sim 3\times 10^{15}$s; one of energy 300 EeV has a
lifetime of about half that Stecker (1968). These values correspond to 
linear propagation distances of $\sim$ 30 and 15 Mpc respectively. 
Even shorter lifetimes were calculated for Fe nuclei, 
based on photodisintegration off the IBR (PSB).
Recent estimates of the lifetimes of UHCR {\it $\gamma$-rays} against 
electron-positron pair production interactions with background radio photons
give values below $10^{15}$s (\cite{pro96}).
Within such distances, it is 
difficult to find candidate sources for UHCRs of such energies.

In this paper, we reexamine a part of the propagation problem by presenting
the results of a new calculation of the photodisintegration of UHCR 
{\it nuclei} through the CBR and IBR in intergalactic space. 
In order to do this,
we have made use of a new determination of the IBR
based on empirical data, primarily from IRAS galaxies, recently calculated by
Malkan \& Stecker (1998).\footnotemark{}\footnotetext{In a recent paper 
(\cite{ste98}), 
one of us considered such
photodisintegration interactions, using the 
new estimate of the extragalactic IR spectral energy distribution by
(\cite{mal98}), which yielded an extragalactic IR photon density 
even lower than the LIR estimate used in PSB. That paper, however, 
neglected the fact that the lower IR flux used in the new calculation 
would imply that interactions with CBR photons would now be dominant 
above an energy of roughly 130 EeV, rather than 300 EeV ({\it cf.} Fig. 8 in
PSB). Thus, as pointed out by Epele \& Roulet (1998), it is 
important to include interactions with CBR photons in the calculation.}

They calculated the intensity and 
spectral energy distribution (SED) of the IBR based on empirical data, some 
of which was obtained for almost 3000 IRAS galaxies. It is these sources 
which produce the IBR. The data used for the new IBR calculation included 
(1) the luminosity dependent SEDs of these galaxies, (2) the 60 $\mu$m 
luminosity function for these galaxies, and (3) the redshift distribution of 
these galaxies. The magnitude of the IBR flux derived by Malkan \& Stecker 
(1998) is is considerably lower than that used in PSB
in their extensive examination of the photodisintegration of UHCR
nuclei. 

A search for absorption in the high energy $\gamma$-ray spectra of 
extragalactic sources can also be used to help determine the value of the 
IBR or to place constraints on the magnitude of its flux (\cite{ste92}).
The observed lack of strong absorption in the $\gamma$-ray spectra of 
the active galaxies Mrk 421 (\cite{mce97}) and Mrk 501 (\cite{aha97}) up to an
energy greater than $\sim$ 5-10 TeV is consistent with the new, lower value 
for the IBR used here (\cite{ste97,ste98a,sta98,bil98}).

The SED calculated by Malkan \& Stecker (1998) agrees with direct estimates
of the far infrared background obtained from the {\it COBE/FIRAS} observations
(\cite{pug96,fix97,fix98}). Recent fluxes reported from 
{\it COBE/DIRBE} obervations at 140 and 240 $\mu$m (\cite{hau98}) 
are roughly a factor of 2 higher than the Malkan \& Stecker (1998) 
predictions, 
but are consistent with them if one considers the systematic 
uncertainties in the observational results (\cite{dwe98}).

In justifying our reexamination of the photodisintegration problem using the
new IBR SED, we point out that it may reasonable to expect that the highest
energy cosmic rays may be nuclei. This is because the maximum energy to which
a particle can be accelerated in a source of a given size and magnetic field
strength is proportional to its charge, $Ze$. That charge is 26 times larger 
for Fe than it is for protons. Although some composition measurements in
the energy range 0.1-10 EeV appear to indicate a transition from heavier to
lighter nuclei with increased energy (\cite{gai93}), this and other data 
appear to be consistent with a ``mixed'' composition of both protons and
heavier nuclei (\cite{hay95,daw98}). 
In any case, at the ``lower'' energies for which composition measurements 
have been attempted, most of the cosmic rays may be galactic in origin.

%In fact, for propagation times on the order of $10^{15}$ s, where the
%photodisintegration cutoff energy is of the order of 200 EeV (see
%Figure 15 of PSB).
%the effect of the CBR on the propagation of particles above this cutoff 
%energy is so important that lowering the infrared flux below the LIR value 
%used in PSB makes almost no effect on the position 
%of the cutoff. It is only in cases of much longer 
%propagation times and correspondingly lower 
%cosmic ray energies, where interactions with infrared photons 
%dominate over interactions with CBR photons, that the effect of lowering the 
%infrared flux results in somewhat higher cutoff energies than those given by
%PSB. However, those cutoff energies, which 
%correspond to propagation distances greater than 100 Mpc, are below 100 EeV.
%Thus, they do not relate to the highest energy events observed.

\section{Calculations}

We have now done a full Monte Carlo calculation 
similar to that presented in PSB, but using the new intergalactic 
infrared spectrum given by Malkan \& Stecker (1998).  
In this new calculation, we have also specifically included the threshold 
energies of the various nuclear species to photodisintegration. The reason
that this is now important is that with the CBR playing a relatively more
important role, interactions of the steeply falling cosmic-ray spectrum
near threshold with the Wien tail of the CBR become important relative to the
much flatter IBR photon spectrum. In fact, as we will show, taking account
of the measured higher threshold energies (as opposed to the artificial value
of 2 MeV taken by PSB) increases the value of the cutoff energy for heavy
UHCR nuclei.

Our intent is to both update and improve the PSB results and also to
determine if the highest energy
CR events seen by the Fly's Eye and Akeno groups are consistent with
their being heavy nuclei that have propagated to us from 
candidate active galactic nuclei (\cite{ste98}). 
The energy loss from photodisintegration
has a much stronger dependence on Lorentz factor than on atomic weight, 
and increases strongly with Lorentz factor $\gamma$ (PSB).  
Therefore, to maximize the
propagation distance for a given total particle energy, $E=\gamma AM$
(where $M$ is the nucleon mass), one takes the largest possible mass
number.  Given the abundances of the elements, this nucleus is Fe.
Therefore (as PSB have done) we chose to examine the propagation history 
of nuclei originating as $^{56}$Fe, as this nuclide offers the best chance 
for providing a conventional explanation for the UHCR events.

\subsection{Cross Sections}

The nuclear photodisintegration process is dominated by the giant dipole
resonance (GDR), which peaks in the \gray\ energy range of 10 to 30 MeV
(nuclear rest frame).  Experimental data are generally consistent with
a two-step process: photoabsorption by the nucleus to form a compound
state, followed by a statistical decay process involving the emission of
one or more nucleons from the nucleus (\cite{lev60}). The photoabsorption 
cross section roughly obeys a Thomas-Reiche-Kuhn (TRK) sum rule, {\it viz}.,
\begin{equation}
\Sigma_{\rm d}\equiv\int_{0}^{\infty}\sigma(\epsilon)\,d\epsilon =
\frac{2\pi^{2}e^{2}\hbar}{Mc}\frac{NZ}{A}=60\frac{NZ}{A}
\mbox{mb-MeV},
\end{equation}
where $A$ is the mass number, $Z$ is the nuclear charge,
$N=A-Z$, $M$ is the nucleon mass and $\epsilon$ is the photon energy
in the rest system of the nucleus.
The TRK sum rule is not exact, however, owing to the presence
of nuclear exchange forces (\cite{fee36}), as can be seen in Table 1 
of PSB (which is based on the data tabulations of Fuller, Gerstenberg,
Vander Molen, and Dunn (1973) and related material supplied by E. Fuller.)

PSB approximated the GDR cross section
for a given nuclide ($Z,A$) with the parameterization
\begin{equation}
\sigma_{i}(\epsilon)=\left\{\begin{array}{l@{\quad}l}
\xi_{i}\Sigma_{\rm d}
W_{i}^{-1}e^{-2(\epsilon-\epsilon_{p,i})^{2}/\Delta_{i}^{2}}
\Theta_{+}(\epsilon_{\rm thr})\Theta_{-}
(\epsilon_{1}), & \epsilon_{\rm thr}\le \epsilon 
\le \epsilon_{1},\quad i=1,2 \\
\zeta\Sigma_{\rm d}
\Theta_{+}(\epsilon_{\rm max})\Theta_{-}(\epsilon_{1})/
(\epsilon_{\rm max}-\epsilon_{1}), &  \epsilon_{1}< \epsilon \le
\epsilon_{\rm max} \\
0, & \epsilon> \epsilon_{\rm max} 
\end{array} \right.
\end{equation}
where the $Z$ and $A$ dependence of the width $\Delta_{i}$, the peak energy
$\epsilon_{p,i}$, and the dimensionless integrated cross sections $\xi_{i}$
and $\zeta$ are understood, and are listed in Table 1 of PSB.
$\Theta_{+}(x)$ and
$\Theta_{-}(x)$ are the Heaviside step functions, and
the normalization constants $W_{i}$
are given by
\begin{equation}
W_{i}=\Delta_{i}\sqrt{\frac{\pi}{8}}\left[
\mbox{erf}\left(\frac{\epsilon_{\rm max}-\epsilon_{p,i}}
{\Delta_{i}/\sqrt{2}}\right)+
\mbox{erf}\left(\frac{\epsilon_{p,i}-\epsilon_{1}}
{\Delta_{i}/\sqrt{2}}\right)\right].
\end{equation}
The index $i$ takes the values 1 or 2, corresponding
to single or double nucleon emission during the photodisintegration reaction.
For energies $\epsilon< \epsilon_{1}=30$ MeV, 
the measured cross sections are dominated
by single [$(\gamma,n)$ or $(\gamma,p)$] or double [$(\gamma,2n)$, 
$(\gamma,np)$, $(\gamma,2p)$] nucleon loss, since the threshold energies for
the emission of larger numbers of nucleons are close to or exceed 
$\epsilon_{1}$ (\cite{for87}).
From $\epsilon_{1}$ to $\epsilon_{\rm max}=150$ MeV, the cross
section is approximated as flat, normalized so that the integrated cross section
matches experimental values.  The probability of emission of $k$ nucleons in
this region is given by a distribution function which is independent of the
\gray\ energy (see Table 2 of PSB).
Above $\epsilon_{\rm max}$, detailed cross section data are
more scarce; we follow PSB
by approximating this smaller
residual cross section by zero. Interactions 
with photons of energy greater than $\epsilon_{max}$ will have a negligible 
contribution to 
the photodisintegration process for UHCR energies below 1000 EeV owing to the 
fact that the density of the background photons seen by the UHCR near the
peak of the GDR cross section falls rapidly with energy 
(exponentially along the Wien tail of the CBR, and roughly as 
$\epsilon^{-2}$ in IR-optical region), along with the fact that the
photodissociation cross section is about two orders of magnitude lower
in the \gray\ energy region from $\epsilon_{\rm max}$ to $\sim$1 GeV than
at the GDR peak (\cite{jon73,jon77}). We have verified by numerical
tests that interactions with photons of energy greater than $\epsilon_{max}$ 
indeed have a negligible effect on our calculation.

In the PSB calculation, the GDR threshold energies were taken to be
$\epsilon_{\rm thr}=2.0$ MeV for {\it all} reaction channels.  This value
is far smaller than the true thresholds; single-nucleon emission has a
typical threshold of $\sim10$ MeV, while the double-nucleon emission
energy threshold is typically $\sim 20$ MeV.  Table 1 lists the
energy thresholds for the $(\gamma,n)$, $(\gamma,p)$, $(\gamma,2n)$,
$(\gamma,np)$, $(\gamma,2p)$, and $(\gamma,\alpha)$ 
channels for each nuclide in the $^{56}$Fe
decay chain (taken from Forkman \& Petersson, 1987).  
Owing to the increased
importance of the CBR relative to the IBR that follows from the new, lower
IBR estimates, and the presence of the Wien tail that exponentially increases
the target photon density at the GDR threshold (see Figure 1), 
increasing this threshold
energy may significantly lengthen the propagation distance of a highly
relativistic nucleus.  Therefore we have used the measured threshold
energies for a subset of the reaction channels that are given in
Table 1.

Unfortunately, photodisintegration cross section data are incomplete.
For many reaction channels, $\sigma(\epsilon)$ data do not exist. Also,
integrated cross section strengths are not available for all of the 
exclusive channels.
The most complete compilation of the world's GDR cross section data 
exists in the 15 volumes of Fuller \& Gerstenberg (1983).  In these volumes
GDR cross section data for $^{56}$Fe, for example, 
are given only for the $(\gamma,pX)$
channel and the inverse channels $(\alpha,\gamma)$ and $(p,\gamma)$.  

One cannot perform the Monte Carlo calculations in such
a way as to distinguish a $(\gamma,n)$ interaction from a $(\gamma,p)$
reaction when the relative branching ratios are not known.  Instead, we
separately consider only single-nucleon emission (as a single channel) and
double-nucleon emission (as a single channel) in the GDR region up to 
$\epsilon_{1}$. We
take the conservative approach of choosing the energy threshold for single-nucleon
emission to be the {\it lowest} of the two for $(\gamma,n)$ and $(\gamma,p)$
for each nuclide in the decay chain, with a similar choice for the 
two-nucleon emission channel.
Along most of the decay chain from
$^{56}$Fe to $^{1}$H there is only one stable isotope for a given mass $A$.
Since the radioactive decay time to the line of stability is less than the 
one-nucleon photodisintegration loss time for all but three unstable
nuclei, $^{53}$Mn, $^{26}$Al, and $^{10}$Be, we assume decay has brought the
daughter nucleus to the line of stability before the next photon collision, so
that for any given mass $A$ there is a unique charge $Z$.  This clearly
is not the case for mass values $A$ = 54, 50, 48, 46, 40, and 36, where
more than one stable isotope exists, but the absence of cross section data
for each of these isotopes makes this a moot point.

We also note that although the thresholds
are lowest for $\alpha$ emission (due to the $\alpha$'s large binding energy),
the integrated cross section for $\alpha$ emission from $^{56}$Fe, for
example, is over two orders of magnitude lower than the $\Sigma_{\rm d}$ value
for that nuclide
(Skopic, Asai, \& Murphy, 1980; see also Fuller \& Gerstenberg, 1983).
We therefore neglect
the $\alpha$ emission channels entirely in our calculation.

\subsection{Reaction Rates}

For a UHCR nucleus with Lorentz factor $\gamma=E/AM$ propagating
through an isotropic soft photon background with differential number
density $n(\epsilon)$, the photodisintegration
rate $R$ (lab frame) is given by (Stecker 1969)
\begin{equation}\label{rate.eq}
R=\frac{1}{2}\int_{0}^{\infty}d\epsilon\,\frac{n(\epsilon)}{\gamma^{2}\epsilon^{2}}
\int_{0}^{2\gamma\epsilon}d\epsilon^{\prime}\,
\epsilon^{\prime}\sigma(\epsilon^{\prime}),
\end{equation}
where $\sigma$ is the total cross section, summed over the number of
emitted nucleons.  

In our calculations we construct the soft photon
background by summing three components: (1) the 
$T=2.73$K cosmic background radiation (CBR)
from lab frame energies of 
$\epsilon=2.0\times 10^{-6}$ eV to $4.0\times 10^{-3}$ eV, 
(2) the infrared background radiation (IBR) estimated by Malkan
\& Stecker (1998)
from $\epsilon=3.0\times 10^{-3}$ eV to 0.33 eV
(using the two ``best estimates'' shown by heavy lines
in their Figure 2, which we
denote as the ``high IBR'' and ``low IBR'' cases),
and (3) the optical to UV
diffuse, extragalactic photon background estimated by Salamon \& Stecker
(1998) (taking their no-metallicity-correction case)
from $\epsilon=0.33$ eV to 13 eV.  Figure 1 shows the SED of the entire
low energy background radiation.  A salient feature of this figure is the
roughly constant energy flux down to the dramatic rise of the Wien tail of
the CBR.  For UHCR nuclei with Lorentz factors large enough so that the CBR
photons are above photodisintegration threshold, most of the reaction rate
$R$ is dominated by collisions with CBR photons.  
This is particularly true
with the new, smaller IR photon background levels, compared to those
used by PSB.

\begin{figure}[h]
\label{fig1}
\epsfysize=3in
\epsfbox{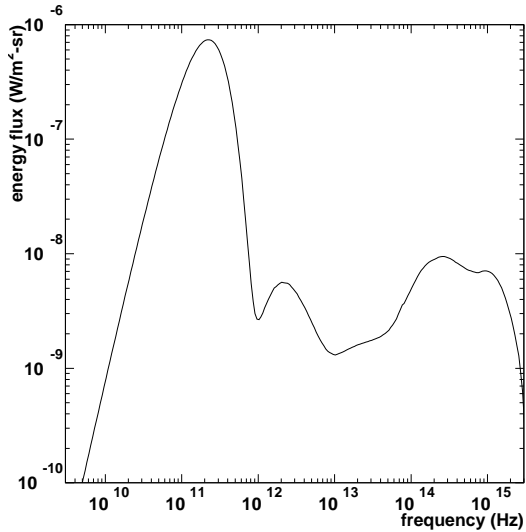}
\caption{Energy flux spectrum (SED) of diffuse, intergalactic soft photon
background.}
\end{figure}

The energy loss time $\tau$ is defined as 
\begin{equation}
\tau^{-1}\equiv\frac{1}{E}\frac{dE}{dt}=
\frac{1}{\gamma}\frac{d\gamma}{dt}+\frac{R_{1}}{A}+\frac{2R_{2}}{A}+
\frac{\left<\Delta A\right>R_{k}}{A},
\end{equation}
where $R_{1}$, $R_{2}$, and $R_{k}$ are the reaction rates for one- and
two-nucleon emission, and $R_{k}$ is the reaction rate for 
$\left<\Delta A\right> >2$ nucleon loss.  The reduction in $\gamma$ comes
from two effects: nuclear energy loss due to electron-positron pair
production off the CBR background, and the \gray\ momentum absorbed by
the nucleus during the formation of the excited compound nuclear state
that preceeds nucleon emission. This latter effect is much smaller 
(of order $\sim10^{-2}$)
than the energy loss from nucleon emission and will therefore be neglected.
For the former mechanism, we use
the results given in Figure 3 of Blumenthal (1970), which gives the loss
rate for relativistic nuclei off the CBR calculated in the first Born
approximation. (We note that the Coulomb corrections to the Born approximation
(\cite{dav54,jau}) have a negligible effect on the pair production loss rate
for ultrarelativistic Fe nuclei.)

Figure 2 shows the energy loss rates due to single-nucleon, double-nucleon,
and pair production processes for $^{56}$Fe as a function of energy,
along with the total energy loss rate.  Also shown is the total energy
loss rate when the photodisintegration thresholds $\epsilon_{\rm thr}$ are
all set to 2 MeV; this indicates the effect of incorporating more realistic
threshold energies in the Monte Carlo calculation, compared to those of PSB.

\begin{figure}[h]
\label{fig2}
\epsfysize=3in
\epsfbox{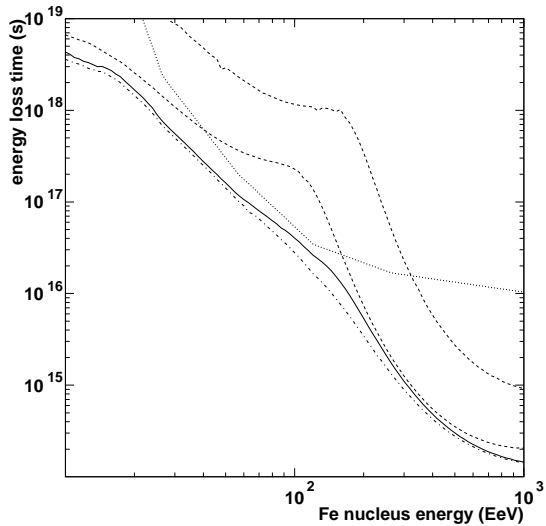}
\caption{Energy loss times for single-nucleon (lower dashed line), 
double-nucleon (upper dashed line), and pair
production processes (dotted line), and their total (solid line).
The dash-dotted line shows the energy loss time using $E_{\rm thr}=2$
MeV for all photodisintegration rates, as was done in PSB.}
\end{figure}

\subsection{Results}

Figure 3 shows the spectra of UHCR nuclei which started out as $^{56}$Fe after 
propagating linear distances
from 1 to 1000 Mpc (corresponding to lifetimes between $10^{14}$ and 
$10^{17}$ s), assuming a source differential energy spectrum
$\propto E^{-3}$ over the energy interval 10 to 1000 EeV.  The effect of
the CBR ``wall'' at the Wien side of the 2.73K blackbody spectrum is to 
produce a sharp cutoff at a rather well defined  energy, $E_{c}$.

\begin{figure}[ht]
\label{fig3}
\epsfysize=3in
\epsfbox{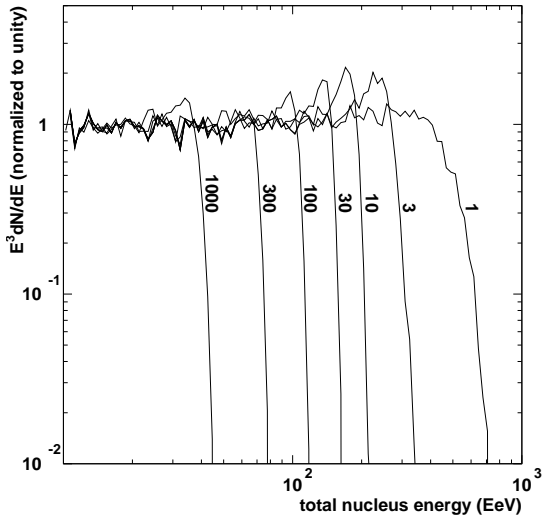}
\caption{The differential spectra of UHCR nuclei ($Z > 1$) which started 
out as $^{56}$Fe after propagation over
the distance (in Mpc) indicated for each curve.  The initial, source
spectrum is a power-law, $E^{-3}$, over the interval 10 to 1000 EeV.}
\end{figure}

Figure 4 shows the cutoff energy 
as a function of propagation time, where $E_{\rm c}$ is
defined as the CR energy at which the propagated differential flux is $1/e$
that of the unpropagated flux. For comparison, the cutoff energies calculated
by PSB are also shown. It can be seen that (except for energies above 
$\sim 200$ EeV) for a given energy, the propagation
time increases by a substantial factor over that calculated by PSB, who 
assumed larger IBR fluxes and a 2 MeV threshold for all photodisintegration 
interactions. We also note that our new results do not differ significantly
for the two SEDs adopted from Malkan \& Stecker (1998) and, except for the
longest propagation times, they do not differ significantly from the no-IBR
case shown in the Figure. This is because the new values obtained for the IBR
are so low.

\begin{figure}[ht]
\label{fig4}
\epsfysize=3in
\epsfbox{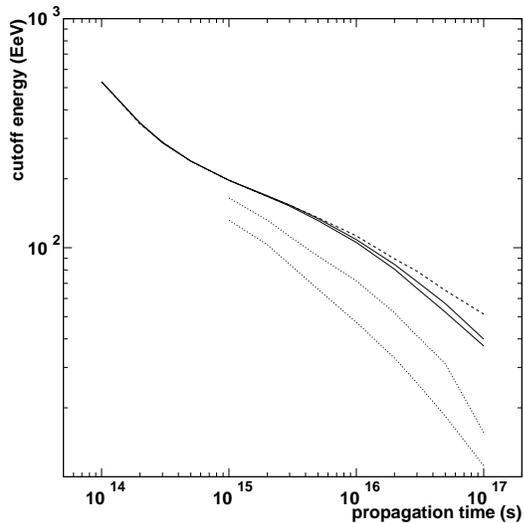}
\caption{Cutoff energy versus propagation time for UHCR nuclei starting out as 
$^{56}$Fe. The solid lines are for our new calculations using the 
two IBR SEDs 
from Malkan \& Stecker (1998) and the higher threshold energies. The dashed 
line is calculated using only the
CBR (no IBR). The dotted lines are the results from Fig. 16 of PSB and are
shown for comparison. The lower the IBR, the higher the cutoff energy curve.}
\end{figure}

\section{Discussion and Conclusions}

As can be seen from Figure 4, our use of the new, lower values for the 
intergalactic infrared photon flux, together with the explicit inclusion of the
measured threshold energies for photodisintegration of the individual nuclides
involved in the calculation, has the effect of 
increasing the cutoff energy (for a given propagation time) of heavy 
UHCR nuclei over that originally calculated by PSB. This increase may have 
significant consequences for understanding the origin of the highest energy
cosmic ray air shower events.

Stanev \etal (1995) and Biermann (1998) have examined the arrival directions
of the highest energy events. They point out that
the $\sim$200 EeV event is within 10$^\circ$ of the direction of the 
strong radio galaxy NGC315.
NGC315 lies at a distance of only $\sim$ 60 Mpc 
from us. For that distance, our results indicate that heavy nuclei would
have a cutoff energy of $\sim$ 130 EeV, which may be within the uncertainty
in the energy determination for this event.  
The $\sim$300 EeV event is within 12$^\circ$ of the strong radio galaxy 3C134. 
The distance to 3C134 is unknown because its location behind a dense molecular cloud 
in our Galaxy obscures the spectral lines required for a redshift measurement.
It may therefore be possible that {\it either} 
cosmic ray protons (Stecker 1968) {\it or} heavy nuclei originated in 
these sources and produced these highest energy air shower events.

\begin{deluxetable}{rrrrrrrr}
\tablecolumns{8}
\tablecaption{Photodisintegration energy thresholds (in MeV) for one-nucleon, 
two-nucleon, and $\alpha$ emission for all isotopes in the decay chain
of $^{56}$Fe \label{tab1}}
\tablehead{\colhead{Z} &  \colhead{A} & \colhead{$(\gamma,n)$}
& \colhead{$(\gamma,p)$} & \colhead{$(\gamma,2n)$} & \colhead{$(\gamma,np)$}
& \colhead{$(\gamma,2p)$} & \colhead{$(\gamma,\alpha)$}}
\startdata
26 & 56 & 11.2 & 10.2 & 20.5 & 20.4 & 18.3 & 7.6 \nl
26 & 54 & 13.4 & 8.9 & 24.1 & 20.9 & 15.4 & 8.4   \nl
25 & 55 & 10.2 & 8.1 & 19.2 & 17.8 & 20.4 & 7.9   \nl
24 & 54 & 9.7 & 12.4 & 17.7 & 20.9 & 22.0 & 7.9   \nl
24 & 53 & 7.9 & 11.1 & 20.0 & 18.4 & 20.1 & 9.1   \nl
24 & 52 & 12.0 & 10.5 & 21.3 & 21.6 & 18.6 & 9.4  \nl
24 & 50 & 13.0 & 9.6 & 23.6 & 21.1 & 16.3 & 8.6   \nl
23 & 51 & 11.1 & 8.1 & 20.4 & 19.0 & 20.2 & 10.3  \nl
23 & 50 & 9.3 & 7.9 & 20.9 & 16.1 & 19.3 & 9.9    \nl
22 & 50 & 10.9 & 12.2 & 19.1 & 22.3 & 21.8 & 10.7 \nl
22 & 49 & 8.1 & 11.4 & 19.8 & 19.6 & 20.8 & 10.2  \nl
22 & 48 & 11.6 & 11.4 & 20.5 & 22.1 & 19.9 & 9.4  \nl
22 & 47 & 8.9 & 10.5 & 22.1 & 19.2 & 18.7 & 9.0   \nl
22 & 46 & 13.2 & 10.3 & 22.7 & 21.7 & 17.2 & 8.0  \nl
21 & 45 & 11.3 & 6.9 & 21.0 & 18.0 & 19.1 & 7.9   \nl
20 & 48 & 9.9 & 15.8 & 17.2 & 24.2 & 29.1 & 14.4  \nl
20 & 46 & 10.4 & 13.8 & 17.8 & 22.7 & 22.7 & 11.1 \nl
20 & 44 & 11.1 & 12.2 & 19.1 & 21.8 & 21.6 & 8.8  \nl
20 & 43 & 7.9 & 10.7 & 19.4 & 18.2 & 19.9 & 7.6   \nl
20 & 42 & 11.5 & 10.3 & 19.8 & 20.4 & 18.1 & 6.2  \nl
20 & 40 & 15.6 & 8.3 & 29.0 & 21.4 & 14.7 & 7.0  \nl
19 & 41 & 10.1 & 7.8 & 17.9 & 17.7 & 20.3 & 6.2   \nl
19 & 40 & 7.8 & 7.6 & 20.9 & 14.2 & 18.3 & 6.4    \nl
19 & 39 & 13.1 & 6.4 & 25.2 & 18.2 & 16.6 & 7.2   \nl
18 & 40 & 9.9 & 12.5 & 16.5 & 20.6 & 22.8 & 6.8   \nl
18 & 38 & 11.8 & 10.2 & 20.6 & 20.6 & 18.6 & 7.2  \nl
18 & 36 & 15.3 & 8.5 & 28.0 & 21.2 & 14.9 & 6.6   \nl
17 & 37 & 10.3 & 8.4 & 18.9 & 18.3 & 21.4 & 7.8   \nl
17 & 35 & 12.6 & 6.4 & 24.2 & 17.8 & 17.3 & 7.0   \nl
16 & 36 & 9.9 & 13.0 & 16.9 & 21.5 & 25.0 & 9.0   \nl
16 & 34 & 11.4 & 10.9 & 20.1 & 21.0 & 20.4 & 7.9   \nl
16 & 33 & 8.6 & 9.6 & 23.7 & 17.5 & 18.2 & 7.1    \nl
16 & 32 & 15.0 & 8.9 & 28.1 & 21.2 & 16.2 & 6.9   \nl
15 & 31 & 12.3 & 7.3 & 23.6 & 17.9 & 20.8 & 9.7   \nl
14 & 30 & 10.6 & 13.5 & 19.1 & 22.9 & 24.0 & 10.6 \nl
14 & 29 & 8.5 & 12.3 & 25.7 & 20.1 & 21.9 & 11.1  \nl
14 & 28 & 17.2 & 11.6 & 30.5 & 24.6 & 19.9 & 10.0 \nl
13 & 27 & 13.1 & 8.3 & 24.4 & 19.4 & 22.4 & 10.1  \nl
12 & 26 & 11.1 & 14.1 & 18.4 & 23.2 & 24.8 & 10.6 \nl
12 & 25 & 7.3 & 12.1 & 23.9 & 19.0 & 22.6 & 9.9   \nl
12 & 24 & 16.5 & 11.7 & 29.7 & 24.1 & 20.5 & 9.2   \nl
11 & 23 & 12.4 & 8.8 & 23.5 & 19.2 & 24.1 & 10.5  \nl
10 & 22 & 10.4 & 15.3 & 17.1 & 23.4 & 26.4 & 9.7  \nl
10 & 21 & 6.8 & 13.0 & 23.6 & 19.6 & 23.6 & 7.3   \nl
10 & 20 & 16.9 & 12.8 & 28.5 & 23.3 & 20.8 & 4.7  \nl
9 & 19 & 10.4 & 8.0 & 19.6 & 16.0 & 23.9 & 4.0   \nl
8 & 18 & 8.0 & 15.9 & 12.2 & 21.8 & 29.1 & 6.2   \nl
8 & 17 & 4.1 & 13.8 & 19.8 & 16.3 & 25.3 & 6.4   \nl
8 & 16 & 15.7 & 12.1 & 28.9 & 23.0 & 22.3 & 7.2  \nl
7 & 15 & 10.8 & 10.2 & 21.4 & 18.4 & 31.0 & 11.0 \nl
7 & 14 & 10.6 & 7.6 & 30.6 & 12.5 & 25.1 & 11.6   \nl
6 & 13 & 4.9 & 17.5 & 23.7 & 20.9 & 31.6 & 10.6  \nl
6 & 12 & 18.7 & 16.0 & 31.8 & 27.4 & 27.2 & 7.4  \nl
5 & 11 & 11.5 & 11.2 & 19.9 & 18.0 & 30.9 & 8.7  \nl
5 & 10 & 8.4 & 6.6 & 27.0 & 8.3 & 23.5 & 4.5     \nl
4 & 9 & 1.7 & 16.9 & 20.6 & 18.9 & 29.3 & 2.5   \nl
3 & 7 & 7.3 & 10.0 & 12.9 & 11.8 & 33.5 & 2.5   \nl
3 & 6 & 5.7 & 4.6 & 27.2 & 3.7 & 26.4 & 1.5     \nl
2 & 4 & 20.6 & 19.8 & 28.3 & 26.1 & \nodata & \nodata \nl
2 & 3 & 7.7 & 5.5 & \nodata & \nodata & \nodata & \nodata \nl
1 & 2 & 2.2 & \nodata & \nodata & \nodata & \nodata & \nodata \nl
\enddata
\end{deluxetable}

\end{document}